\documentclass[a4paper,12pt]{article}

\usepackage{amsmath,amssymb,mathrsfs}
\usepackage[dvips]{graphicx}
\usepackage{amsthm,bm,
	color}
\usepackage[color]{showkeys}
\usepackage{enumerate}
\makeatletter
\def\figcaption{\def\@captype{figure}\caption}
\makeatother

\makeatletter

\@addtoreset{equation}{section}
\renewcommand{\p@enumi}{C.}
\makeatother
\usepackage{bbm}
\usepackage{comment}
\usepackage{amsthm}
\usepackage{bm}
\usepackage{color}
\usepackage{cite}

\def\i<#1>{\langle #1 \rangle}
\def\l<#1>{\left\langle #1 \right\rangle}

\makeatother

\newtheorem{theorem}{Theorem}[section]
\newtheorem{definition}[theorem]{Definition}
\newtheorem{proposition}[theorem]{Proposition}
\newtheorem{lemma}[theorem]{Lemma}

\newtheorem{remark}[theorem]{Remark}

\makeatletter

\def\@thesis{}
\def\id#1{\def\@id{#1}}
\def\department#1{\def\@department{#1}}

\def\@maketitle{
	\begin{center}
		{\LARGE \@title \par}%
		\vspace{5mm}
		{\Large \@author \par}%
		\vspace{5mm}
		
	\end{center}
	\par\vskip 1.5em
}

\makeatother

\title{{\bf Eigenvalues and threshold rezonances of a two-dimensional split-step quantum walk with strong shift}}
\author{Toru Fuda\footnote{
		School of Science and Engineering, Kokushikan University, Setagaya, Tokyo 154-8515, Japan, 
		\\E-mail: fudat@kokushikan.ac.jp},\ \ 
	Daiju Funakawa\footnote{Department of Electronics and Information Engineering, Hokkai-Gakuen University, Sapporo 062-8605,
		Japan.
		\\E-mail: funakawa@hgu.jp},\ \ 
	Satoshi Sasayama\footnote{Department of Information Media, Hokkaido Information University, Nishi-Nopporo 59-2, Ebetsu, Hokkaido 069-8585, Japan,
		\\E-mail:sasayama@do-johodai.ac.jp},\ \ 
	Akito Suzuki\footnote{Division of Mathematics and Physics, Faculty of Engineering, Shinshu University, Wakasato, Nagano
		\\380-8553, Japan, E-mail: akito@shinshu-u.ac.jp}}

\begin{document}
	
	\maketitle
	
	\begin{abstract}
		In this paper, we derive sufficient conditions for the localization of two-dimensional split-step quantum walks with a strong shift. For this purpose, we analyze the zero points of the function $f$ introduced by Fuda et.al. (Quantum Inf Process 16(8) 203, 2017) and make these zero points explicit. These zeros provide a concrete representation of the eigenvalues and eigenvectors of the evolution operator, and in particular, clarify where localization occurs. 
		In addition, 
		the eigenvalues obtained here asymptotically approach threshold resonance in special cases.
		We also describe the display of threshold resonances and generalized eigenfunctions.
	\end{abstract}

	\section{Introduction}
	
	Quantum walks have been actively studied in recent years, being introduced to quantum information \cite{LTRSW,S}, quantum simulations \cite{SG,Si}, and topological phases \cite{Ki, KBF}.
	Split-step quantum walks (SSQWs), introduced by Kitagawa et.al. \cite{Ki} for the analysis of topological phases, have been studied in various fields including localized phenomena \cite{FFS,FFS2,FNSS}, weakly convergent limit \cite{FFS3}, index theory \cite{Ma,MST,T} and so on \cite{MC,NOW}. 
	A necessary and sufficient condition for a quantum walk to induce localization is that the time evolution operator has eigenvalues \cite{SeSu}. 
	In particular, non-zero components of eigenvectors are the points at which quantum walk localization occurs.
	
	In the context of the discrete Schr\"{o}dinger equation, a value called resonance is known \cite{ALMM,HL}. 
	If a Hamiltonian satisfies the eigenequations but its eigenfunctions are not square integrable, its eigenvalues are called resonances and its eigenfunctions are called the generalized eigenfunctions corresponding to the resonances.
	The same thing can happen when considering the eigenequations of the time evolution operator described by quantum walks. 
	As mentioned above, it is equivalent for a quantum walk to cause localization and for the time evolution operator of a quantum walk to have eigenvalues, but the eigenfunctions corresponding to those eigenvalues may not be square integrable.
	Therefore, it is natural that the time evolution operator of this quantum walk has resonance.
	Prior work on quantum walk resonances and generalized eigenfunctions can be found in \cite{KKMS,KKMS2,Mo}. 
	
	Resonances for a quantum walk $U$ also apply to stationary measures. 
	If $\|(U^t\Psi _0)(x)\|^2$ is independent with respect to time $t\in \mathbb {Z}_{\geq 0}$, then the measure $\mu (x):=\| (U^t\Psi _0)(x)\|^2$ is called a stationary measure \cite{EKKT19,KK17,KK22, KKK}, where $\Psi _0$ is the initial state. 
	If the initial state $\Psi _0$ is a generalized eigenfunction of $U$, then $\| (U^t\Psi _0)(x)\|^2$ is time independent. 
	Therefore, the research of resonances is linked to the stationary state. 
	
	In fact, generalized eigenfunctions have been used to derive the stationary measure of quantum walks \cite{EKKT19,KK17,KK22, KKK}.
	In \cite{KKK}, the stationary measures corresponding to the generalized eigenfunctions of one-dimensional space-inhomogeneous quantum walk with finite defects are presented. 
	The stationary measures of a multi-dimensional space-homogeneous quantum walk with cycles can be found in \cite{KK17}. 
	The eigenvalues of the quantum walk can be divided into two essentially different parts \cite{SS}. 
	That is, the part derived from birth and the part inherited from the discriminant operator $T$. 
	A multi-dimensional quantum walk with cycles has an infinite number of eigenvectors with finite support, which corresponds to the birth eigenspace. 
	In \cite{KK17}, stationary measures are constructed by the superposition of an infinite number of eigenvectors in the birth eigenspace.
	On the other hand, in this paper, we derive the generalized eigenfunctions of the two-dimensional SSQWs not from the birth eigenspace, but from the inherited eigenspace.
	Thus, the eigenfunction in this paper differs from that in \cite{KK17} in its origins. 
	
	It was shown in \cite{FFS} that in $d$-dimensional SSQWs, if the effect of the shift is sufficiently weak, i.e., 
	transition probability $|q_j|^2$ is sufficiently small, the time evolution operator has eigenvalues in the inherited eigenspace.
	We are interested in whether the time evolution operator has eigenvalues when $|q_j|^2$ is sufficiently large.
	Therefore, in this paper, a detailed analysis is performed under the condition that the shift is extremely strong, i.e, $q_1 = q_2 = 1$ in two-dimensional SSQWs. 
	In our model, eigenvalues and eigenvectors can be specifically displayed. 
	In addition, we derive resonances and corresponding generalized eigenfunctions in a special case. 
	
	The remainder of this paper is organized as follows. 
	Section \ref{sec_2} first introduces the general form of a two-dimensional SSQW with one defect 
	and present prior work on localization \cite{FFS}. 
	While the previous study dealt only with the case where the effect of the shift is weak, 
	this study deals with a model in which the effect of the shift is strong. 
	We introduce conditions for this purpose.
	The main results are presented in Section \ref{sec_3}.
	In Section \ref{sec_3}, we give sufficient conditions for SSQWs to cause localization.
	If that sufficient condition is satisfied, we also show that the time evolution operator has at least four eigenvalues. 
	It also showed where exactly localization is occurring.
	Furthermore, $|\phi _{2,2}|<|\omega _{2,2}|$ is a sufficient condition for localization to occur, but $U$ have resonances when $|\phi _{2,2}|=|\omega _{2,2}|$. 
	In particular, these resonances are boundary values of the essential spectrum of $U$ and are called threshold resonances.

	\section{Model definition and related results}\label{sec_2}
	In this section, we first define a general two-dimensional 
	split-step quantum walks (SSQWs) and briefly introduce related results. 
	For detailed background and discussion, see  \cite{FFS}. 
	Next, we define a more specific model of SSQWs that can be interpreted as an 
	``extremely {\it strong} shift effect''.
	The purpose of this paper is to shed light on a new aspect of SSQWs through the analysis of this specific model. 
	Note that while  \cite{FFS} provides sufficient conditions for the localization of SSQWs that can be interpreted as ``very {\it weak} shift effects'',  the model presented in this paper deviates significantly from it.
	
	\subsection{Definition of two-dimensional split-step quantum walks}
	In what follows, we use ${\mathcal H}$ to denote $\ell^2(\mathbb {Z}^2;\mathbb {C} ^4)$.
	Let $(\bm{p}, \bm{q}) = (p_1, p_2, q_1, q_2)\in \mathbb{R}^2\times \mathbb{C}^2$ satisfy 
	$p_j^2+|q_j|^2=1\ (j=1,2)$
	and use $\{\bm{e}_j\}_{j=1}^2$ to denote the standard basis of $\mathbb{Z}^2$. 
	We set an operator $S_j\ (j=1,2)$ on $\ell^2(\mathbb{Z}^2; \mathbb{C}^2)$ as 
	$\displaystyle
	S_j:=
	\begin{pmatrix}
		p_j&q_jL_j\\
		(q_jL_j)^*&-p_j
	\end{pmatrix},
	$
	where $L_j$ is the $\bm{e}_j$-shift on $\ell^2(\mathbb{Z}^2)$ defined by 
	$(L_jf)(\bm{x})=f(\bm{x}+\bm{e}_j)$ for all $f\in \ell^2(\mathbb{Z}^2)$, i.e., 
	\begin{align*}
		(S_j\psi )({\bm x})=
		\begin{pmatrix}
			p_j\psi _1({\bm x})+q_j\psi _2({\bm x}+{\bm e_j})\\
			q_j^*\psi _1({\bm x}-{\bm e}_j)-p_j\psi _2({\bm x})
		\end{pmatrix}
		\quad \text{for all}\ 
		{\bm x}\in \mathbb{Z}^2,\ \psi =\begin{pmatrix}
			\psi_1\\
			\psi_2
		\end{pmatrix}\in \ell^2(\mathbb{Z}^2;\mathbb{C}^2).
	\end{align*}
	Identifying $\mathcal{H}$ with $\ell^2(\mathbb{Z}^2;\mathbb{C}^2)\oplus \ell^2(\mathbb{Z}^2;\mathbb{C}^2)$, we define the shift operator $S$ on $\mathcal{H}$ as 
	$S=S_1\oplus S_2$, i.e., 
	\begin{align*}
		(S\Psi)(\bm{x}) = 
		\begin{pmatrix}
			(S_1\Psi_1)(\bm{x})\\
			(S_2\Psi_2)(\bm{x})
		\end{pmatrix}
		\quad \text{for all}\ 
		{\bm x}\in \mathbb{Z}^2,\ 
		\Psi =
		\begin{pmatrix}
			\Psi_1\\
			\Psi_2
		\end{pmatrix}\in \mathcal{H}, \
		\Psi_j \in \ell^2(\mathbb{Z}^2;\mathbb{C}^2)\ (j=1,2). 
	\end{align*}
	Then $S$ is self-adjoint and unitary on $\mathcal{H}$ and satisfies $S^2=1$. 
	Next, we define the coin operator. Let $\Omega ={}^t(\omega _{1,1}, \omega _{1,2}, \omega _{2,1}, \omega _{2,2})$ and $\Phi ={}^t(\phi _{1,1}, \phi _{1,2}, \phi _{2,1}, \phi _{2,2})$ be normalized vectors on $\mathbb{C}^4.$
	We define the coin operator $C$ on $\mathcal{H}$ as a multiplication operator by $C({\bm x} )$, i.e.,  
	\begin{align*}
		(C\Psi)(\bm{x}) = C(\bm{x})\Psi(\bm{x})\quad 
		\text{for all} \ \bm{x}\in\mathbb{Z}^2, \ \Psi\in \mathcal{H}. 
	\end{align*}
	In our one-defect model, $C(\bm{x})$ is defined by 
	\begin{align*}
		C(\bm{x}) &= 2|\chi(\bm{x})\rangle\langle \chi(\bm{x})| - 1, \\
		\chi(\bm{x}) &=
		{}^t(\chi_{1,1}(\bm{x}), \chi_{1,2}(\bm{x}), \chi_{2,1}(\bm{x}), \chi_{2,2}(\bm{x}))
		= 
		\begin{cases}
			\Phi 
			, &{\bm x}\in\mathbb{Z}^2\setminus \{\bf 0\},\\
			\Omega 
			, &{\bm x}=\bm{0}.
		\end{cases}
	\end{align*}
	Then $C$ is self-adjoint and unitary on $\mathcal{H}$.
	We define a time evolution operator $U$ on $\mathcal{H}$ as $U=SC$. 
	Since both $S$ and $C$ are self-adjoint and unitary on $\mathcal{H}$, 
	the spectral mapping theorem (SMT) of quantum walks  \cite{SS} can be applied to $U$. 
	
	\subsection{Related results for localization}
	To introduce the SMT results, we provide some definitions. 
	The discriminant operator $T$ on $\ell^2(\mathbb{Z}^2)$ with respect to $U$ is defined as 
	$T=dSd^*$, 
	where $d : \mathcal{H}\to \ell^2(\mathbb{Z}^2)$ is a boundary operator, i.e., 
	\begin{align*}
		(d\Psi)(\bm{x}) = \langle\chi(\bm{x}), \Psi(\bm{x})\rangle_{\mathbb{C}^4}\quad
		\text{for all} \ \bm{x}\in \mathbb{Z}^2, \Psi\in\mathcal{H}.
	\end{align*}
	Furthermore, for a boundary operator $d$, the following holds:
	\begin{align*}
		&(d^*f)(\bm{x}) = \chi(\bm{x})f(\bm{x})\quad \text{for all}\ \bm{x}\in\mathbb{Z}^2, f\in\ell^2(\mathbb{Z}^2),\\
		&d^*d = \bigoplus_{\bm{x}\in\mathbb{Z}^2}|\chi(\bm{x})\rangle\langle\chi(\bm{x})|, \quad dd^* = I_{\ell^2(\mathbb{Z}^2)},\\
		&C = 2d^*d - 1.
	\end{align*}
	
	\begin{definition}[Localization]
		Let $\Psi _0\in {\mathcal H}$ be an initial state with $\| \Psi _0\|=1$. 
		We say that a pair $(U,\Psi _0)$ causes localization when there exists ${\bm x}\in \mathbb{Z}^2$ such that  $\limsup _{t\to \infty }\| (U^t\Psi _0)(\bm{x})\|^2>0$ holds.
	\end{definition}
	
	In the following, sufficient conditions for localization are described. 
	By \cite[Proposition~2.4]{SeSu}, $(U, \Psi _0)$ causes localization if and only if the time evolution operator $U$ has eigenvalue $\mu$ and $\Psi _0$ overlap with the eigenspace $\ker(U-\mu)$. 
	Therefore, for localization to occur, $U$ must have an eigenvalue. 
	We briefly present known results on the sufficient conditions for $U$ to have eigenvalues. 
	Among the results obtained by SMT, the following are of particular interest in this paper: 
	\begin{align}
		\{ e^{\pm i\arccos \lambda }\mid \lambda \in \sigma _{\rm p}(T)\}\subset \sigma _{\rm p}(U),\quad \{ e^{\pm i\arccos \lambda }\mid \lambda \in \sigma _{\rm ess}(T)\}= \sigma _{\rm ess}(U).\label{SMT}
	\end{align}
	From this result, sufficient conditions for localization can be obtained by analyzing the eigenvalues of $T$.
	The following is a preparation for analyzing the eigenvalues of $T$. 
	We define the bounded operator $T_0$ on $\ell^2(\mathbb{Z}^2)$ as follows:
	\begin{align}
		&T_0=a_{\Phi }({\bm p})+\sum _{j=1}^2\left( q_j\phi _{j,1}^*\phi _{j,2}L_j+q_j^*\phi _{j,1}\phi _{j,2}^*L_j^* \right), \label{eq:t}\\
		&a_{\Phi }({\bm p})=\sum _{j=1}^2p_j\left(|\phi _{j,1}|^2-|\phi _{j,2}|^2\right), \quad
		a_{\Omega }({\bm p})=\sum _{j=1}^2p_j\left(|\omega _{j,1}|^2-|\omega _{j,2}|^2\right)\nonumber.
	\end{align}
	By \cite[Lemma 3.3]{FFS}, we find 
	\begin{align}
		\sigma_{\rm ess}(T)=\sigma (T_0)=
		\left[-2\sum _{j=1}^2|q_j\phi _{j,1}\phi _{j,2}|+a_{\Phi }({\bm p}),2\sum _{j=1}^2|q_j\phi _{j,1}\phi _{j,2}|+a_{\Phi }({\bm p})\right]. \label{eq:ess}
	\end{align}
	Let $\bm{1}_A$ be the characteristic function of a set $A$. 
	Assuming $[-1, 1]\setminus \sigma (T_0)\neq \emptyset$, we can define the function $f$ on $[-1, 1]\setminus \sigma (T_0)$ as
	\begin{align*}
		f(\lambda ) = \lambda +\langle \varphi _{\bm q},\psi _{\lambda }\rangle_{\ell^2(\mathbb{Z}^2)}, 
	\end{align*}
	where 
	\begin{align}\label{eq:phipsi}
		\varphi _{\bm q}=\sum _{j=1}^2\left( q_j\omega _{j,2}\phi _{j,1}^*{\bm 1}_{\{-{\bm e}_j\}}+q_j^*\omega _{j,1}\phi _{j,2}^*{\bm 1}_{\{ {\bm e}_j\}} \right),\quad 
		\psi _{\lambda }=\left( T_0-\lambda \right)^{-1}\varphi _{\bm q}.
	\end{align}
	By the method of the paper  \cite{FFS}, we obtain the following facts for $f$ and the eigenvalues of $T, U$:
	\begin{align}\label{eigenvalue_corresp}
		f(\lambda )=0 \ \Longrightarrow \ \lambda \in \sigma_{\rm p}(T) \ \Longrightarrow \ e^{\pm i\arccos \lambda }\in \sigma _{\rm p}(U).
	\end{align}
	Note that the eigenvalues $e^{\pm i\arccos \lambda }$ of $U$ are discrete. 
	The next theorem gives a sufficient condition for $U$ to have eigenvalues. 
	\begin{theorem}[ \cite{FFS},\ $d=2$]\label{FFS17main}
		Assume the following conditions: 
		\begin{align}
			&\forall j\in \{1,2\}, \ \phi_{j,1}\omega_{j,2}+\phi_{j,2}\omega_{j,1}=0, \label{FFS17_1}\\
			&\exists l\in \{1,2\}, \ \phi_{l,1}\omega_{l,2}\neq 0, \\
			&\exists \bm{p}_0\in \{-1,1\}^2, \ a_{\Phi}(\bm{p}_0) \neq a_{\Omega}(\bm{p}_0). 
		\end{align}
		Then there exists $\delta > 0$ such that if $(\bm{p}, \bm{q})$ satisfies $p_lq_l\neq 0$ and 
		$\|(\bm{p}, \bm{q}) - (\bm{p}_0, \bm{0})\|_{\mathbb{R}^2\times\mathbb{C}^2}<\delta$, then 
		$f$ have a zero, i.e., there exist eigenvalues of $U$. 
	\end{theorem}
	
	Note that Theorem \ref{FFS17main} is applicable only when the shift is sufficiently weak (i.e., $\bm{q}$ is sufficiently close to $\bm{0}$). In the next subsection, we define SSQWs with a strong shift (i.e., $\bm{p} = \bm{0}$). 
	
	\subsection{SSQWs with strong shift}
	We consider SSQWs in which the shift is so strong that Theorem \ref{FFS17main} cannot be applied. 
	Therefore, we introduce the following conditions.		
	\begin{enumerate}[({C.}1)]
		\item $\omega_{1,2}=\phi_{1,2}=0$ \ and \ 
		$\omega _{2,1}\phi _{2,2}+\omega _{2,2}\phi _{2,1}=0$, \label{c1}
		\item $\omega_{i,j}, \phi_{i,j} \in\mathbb{R}\setminus\{0\}$ \ for \ $(i,j)\in\{(1,1),(2,1),(2,2)\}$, \label{c2}
		\item $p_1=p_2=0$ \ and \ $q_1=q_2=1$.\label{c3} 
	\end{enumerate}

	\begin{remark}
		Condition \eqref{c1} is a special case of condition (\ref{FFS17_1}) in Theorem \ref{FFS17main}. 
		Condition \eqref{c2} is not essential, and the rest of the paper's argument can be developed similarly even if the parameters are not real values. 
		Note that condition \eqref{c3} is a ``strong shift'' condition, and the assumption of Theorem \ref{FFS17main} is not satisfied in the model with this condition imposed.
	\end{remark}
	Hereafter, conditions \eqref{c1}--\eqref{c3} is assumed unless otherwise noted. 
	By conditions \eqref{c1}--\eqref{c3}, $a_{\Phi}(\bm{p}) = a_{\Omega}(\bm{p}) = 0$ holds and 
	the previously defined $T_0$ and $\varphi_{\bm{q}}$ can be rewritten as 
	\begin{align*}
		T_0=\phi _{2,1}\phi _{2,2}(L_2+L_2^*), \quad 
		\varphi _{{\bm q}}=\phi _{2,1}\omega _{2,2}{\bm 1}_{\{-{\bm e}_2\}}+\phi_{2,2}\omega_{2,1}{\bm 1}_{\{{\bm e}_2\}}.
	\end{align*}
	Defining $\Lambda := 2|\phi_{2,1}\phi_{2,2}|$, $[-1,1]\setminus \sigma(T_0)\neq \emptyset$ holds because 
	$\sigma(T_0) = [-\Lambda, \Lambda]$ and 
	\begin{align*}
		\Lambda = 2|\phi_{2,1}\phi_{2,2}| \leq |\phi_{2,1}|^2 + |\phi_{2,2}|^2 < 1.
	\end{align*}
	We set ${\mathbb T}_-:=[-1,-\Lambda )$ and ${\mathbb T}_+:=(\Lambda ,1]$, then the domain of $f$ is ${\mathbb T}_-\cup\mathbb{T}_+$. 
	
	\section{Main result}\label{sec_3}
	In this section, we discuss the main results. 
	The main results consist of two parts: one on localization and one on resonance.
	The models treated in this section are SSQWs with the strong shift defined in the previous section. 
	\subsection{Sufficient conditions for localization of strong shift}
	In the next theorem, we present the necessary and sufficient conditions for $f$ to have zero points.
	
	\begin{theorem}\label{main1}
		Under the conditions \eqref{c1}--\eqref{c3}, the following conditions (1) and (2) are equivalent. 
		\begin{enumerate}[{}(1)]
			\item $f$ has zero points on ${\mathbb T}_-\cup {\mathbb T}_+$. 
			\item $|\phi _{2,2}|<|\omega _{2,2}|$. 
		\end{enumerate}
		In particular, when either (i.e., both) of the above conditions holds, 
		the zeros of $f$ consist of $\lambda _0^+$ and $\lambda _0^-$ given by
		\begin{align*}
			\lambda _0^{\pm}=\pm \frac{2\omega _{2,1}\omega _{2,2}}{\sqrt{2\omega _{2,2}^2/\phi _{2,2}^2-1}}.
		\end{align*}
		Note that $\lambda_0^{+}\neq \lambda_0^{-}$ by condition \eqref{c2}. 
		Furthermore, \eqref{eigenvalue_corresp} of the eigenvalue correspondence indicates that $U$ has at least four eigenvalues 
		$e^{i\arccos \lambda _0^{\pm}}$ \ and \ $e^{-i\arccos \lambda _0^{\pm}}$. 
	\end{theorem}
	\begin{proof}
		First, we calculate $f(\lambda)$. By the Fourier transform from $\ell^2(\mathbb{Z}^2)$ to $L^2([0,2\pi)^2, d\bm{k}/(2\pi)^2)$ and condition \eqref{c1}, we obtain $\hat{\varphi} _{\bm q}({\bm k})=2i\phi _{2,1}\omega _{2,2}\sin k_2$ and $\hat{T}_0({\bm k})=2\phi _{2,1}\phi _{2,2}\cos k_2$ for ${\bm k}=(k_1,k_2)\in [0,2\pi)^2$. 
		We set $a:=\lambda /2\phi _{2,1}\phi _{2,2}$ for $\lambda\in \mathbb{T}_-\cup \mathbb{T}_+$. 
		Note that $|a|>1$. 
		Then we have the following equation:
		\begin{align*}
			\frac{f(\lambda )}{2\phi _{2,1}\phi _{2,2}}&=a+\frac{1}{2\phi _{2,1}\phi _{2,2}}\i<\hat{\varphi} ,(\hat{T}_0(\cdot)-\lambda )^{-1}\hat{\varphi} >\\
			&=a+\frac{1}{4\phi ^2_{2,1}\phi ^2_{2,2}}\int _{[0,2\pi)^2}\frac{d{\bm k}}{(2\pi)^2}\frac{4\phi _{2,1}^2\omega _{2,2}^2\sin ^2k_2}{\cos k_2-a }\\
			&=a+\frac{\omega _{2,2}^2}{\phi ^2_{2,2}}\int _{[0,2\pi)}\frac{dk_2}{(2\pi)}\frac{\sin ^2k_2}{\cos k_2-a }\\
			&=a+\frac{\omega _{2,2}^2}{\phi _{2,2}^2}(-a+{\rm sgn}(a)\sqrt{a^2-1}).
		\end{align*}
		Thus, if there exist zero points $\lambda _0\in \mathbb{T}_-\cup\mathbb{T}_+$ of $f$, i.e., $f(\lambda _0)=0$, then 
		\begin{align}
			&\Lambda = 2|\phi_{2,1}\phi_{2,2}| < |\lambda_0| \leq 1, \label{sol_1}\\
			&\left(\frac{\omega _{2,2}^2}{\phi _{2,2}^2}-1\right)|a| = \frac{\omega _{2,2}^2}{\phi _{2,2}^2}\sqrt{a^2-1} \ (>0). \label{sol_2}
		\end{align}
		Note that condition \eqref{sol_1} and \eqref{sol_2} leads to conditions
		\begin{align}
			&|\phi_{2,2}| \neq |\omega_{2,2}|, \quad 
			2|\omega _{2,1}\omega _{2,2}|\leq \sqrt{2\omega _{2,2}^2/\phi _{2,2}^2-1} \label{sol_1'}\\
			&\text{and} \quad 
			|\phi_{2,2}| < |\omega_{2,2}|, \label{sol_2'}
		\end{align}
		respectively, but \eqref{sol_1'} is fully subsumed by \eqref{sol_2'}. 
		From the above discussion, we obtain the following:
		\begin{align}
			\exists \lambda_0\in \mathbb{T}_-\cup\mathbb{T}_+, \ f(\lambda_0) = 0\ 
			\Longrightarrow \ 
			|\phi_{2,2}| < |\omega_{2,2}|.\label{sol_to_con}
		\end{align}
		The reverse direction of \eqref{sol_to_con} can be easily seen.
		When the above conditions (1) or (and) (2) hold, solving equation \eqref{sol_2} yields
		\begin{align*}
			\lambda _0=\pm \frac{2\omega _{2,1}\omega _{2,2}}{\sqrt{2\omega _{2,2}^2/\phi _{2,2}^2-1}}.
		\end{align*}
	\end{proof}

	\begin{theorem}\label{main2}
		Assume (2) in Theorem \ref{main1} under conditions \eqref{c1}--\eqref{c3}. 
		Let $\mu _0$ be one of the eigenvalues of $U$ corresponding to the zero point $\lambda_0$ of $f$, i.e., 
		$\mu_0 \in \{e^{\pm i\arccos\lambda_0}\}, \ f(\lambda_0)=0$. 
		Then the eigenvector $\Psi _{\mu _0}$ of $U$ corresponding to the eigenvalue $\mu _0$ satisfies for all 
		${\bm x} = (x_1, x_2)\in \mathbb{Z}^2$, 
		\begin{align}\label{eq:main2:1}
			\Psi _{\mu _0}({\bm x})=\begin{pmatrix}
				\chi _{1,1}({\bm x})\left( \psi _{\lambda _0}({\bm x})-{\bm 1}_{\{0\}}({\bm x})\right)\\
				-\mu _0\chi _{1,1}({\bm x}-{\bm e}_1)\left( \psi _{\lambda _0}({\bm x}-{\bm e}_1)-{\bm 1}_{\{0\}}({\bm x}-{\bm e}_1)\right)\\
				\chi _{2,1}({\bm x})\left( \psi _{\lambda _0}({\bm x})-{\bm 1}_{\{0\}}({\bm x})\right)-\mu _0\chi _{2,2}({\bm x}+{\bm e}_2)\left( \psi _{\lambda _0}({\bm x}+{\bm e}_2)-{\bm 1}_{\{0\}}({\bm x}+{\bm e}_2)\right)\\
				\chi _{2,2}({\bm x})\left( \psi _{\lambda _0}({\bm x})-{\bm 1}_{\{0\}}({\bm x})\right)-\mu _0\chi _{2,1}({\bm x}-{\bm e}_2)\left( \psi _{\lambda _0}({\bm x}-{\bm e}_2)-{\bm 1}_{\{0\}}({\bm x}-{\bm e}_2)\right)\\
			\end{pmatrix}, 
		\end{align}
		where 
		\begin{align}
			&\psi _{\lambda _0}({\bm x})=[(T_0-\lambda _0)^{-1}\varphi _{\bm q}]({\bm x})={\bm 1}_{\{x_1=0\}}({\bm x})\int _{[0,2\pi)}\frac{2i\omega _{2,2}\phi _{2,1}\sin k_2}{2\phi _{2,1}\phi _{2,2}\cos k_2-\lambda _0}e^{ik_2x_2}\frac{dk_2}{2\pi}. \label{psilambda}
		\end{align}
	\end{theorem}
	\begin{proof}
		We set $P:=I-|{\bm 1}_{\{ \bf 0 \}}\rangle \langle {\bm 1}_{\{ \bf 0 \}}|$
		and $\psi _{\lambda _0}:=(T_0-\lambda _0)\varphi _{\bm q}$. 
		By conditions \eqref{c1} and \eqref{c2}, $\psi _{\lambda _0}\in {\rm Ran}P\backslash \{0\}$ holds. 
		Recall the discussion in  \cite{FFS} using the Feshbach map $F(\lambda)$. 
		Since $\lambda _0$ is a zero of $f$, 
		$\lambda_0$ is also an eigenvalue of $T$, and $F(\lambda _0)\psi _{\lambda _0}=0$ holds. 
		Furthermore, by a theorem on Feshbach maps in the paper \cite[Theorem II.1]{BFS}, the eigenvector of $T$ corresponding to the eigenvalue $\lambda_0$ can be written as follows: 
		\begin{align}\label{proto_h}
			[P-(P^{\perp}(T-\lambda _0)P^{\perp})^{-1}P^{\perp}TP]\psi _{\lambda _0}. 
		\end{align}
		By \cite[Lemma 4.1 and Proposition 4.3]{FFS}, we have $(P^{\perp}(T-\lambda _0)P^{\perp})^{-1}=-\lambda _0^{-1} P^{\perp}$ and $P^{\perp}TP=|{\bm 1}_{\{0\}}\rangle \langle \varphi _{{\bm q}}|.$ Thus 
		\begin{align*}
			[P-(P^{\perp}(T-\lambda _0)P^{\perp})^{-1}P^{\perp}TP]\psi _{\lambda _0}=\psi _{\lambda _0}+\frac{\langle \varphi _{\bm q},\psi _{\lambda _0}\rangle }{\lambda _0}{\bm 1}_{\{ 0 \}}.
		\end{align*}
		Since $\lambda _0$ is a zero point of $f$, $f(\lambda _0)=\lambda _0+\langle \varphi _{\bm q}, \psi _{\lambda _0}\rangle=0$ holds. Thus we have $[P-(P^{\perp}(T-\lambda )P^{\perp})^{-1}P^{\perp}TP]\psi _{\lambda _0}=\psi _{\lambda _0}-{\bm 1}_{\{0\}}.$
		Next, by  \cite{SS}, the following vector is the eigenvector of $U$ corresponding to the eigenvalue 
		$\mu_0 \in \{e^{\pm i\arccos\lambda_0}\}$: 
		\begin{align*}
			\Psi _{\mu _0}=(1-\mu _0S)d^*(\psi _{\lambda _0}-{\bm 1}_{\{{\bf 0}\}}).
		\end{align*}
		By direct calculation, we obtain \eqref{eq:main2:1} and \eqref{psilambda}. 
		Note that \eqref{psilambda} is obtained by Fourier transform and inverse Fourier transform.
	\end{proof}
	
	\begin{remark}
		Theorem \ref{main1} shows that SSQWs with the strong shift are localized as in  \cite{FFS}. 
		Moreover, we see that the localized location ${\bm x}=(x_1,x_2)$ always satisfies $x_1=0$ or $x_1 = 1$ by Theorem \ref{main2}. 
		The cause is condition $\omega_{1,2}=\phi _{1,2}=0$ in \eqref{c1}. 
		If this condition is removed, localization is expected to occur even for $x_1 \neq 0$ and $x_1 \neq 1$. 
	\end{remark}
	
	\subsection{Threshold resonances and generalized eigenfunction}
	In this subsection, we consider the case of $|\phi _{2,2}|=|\omega _{2,2}|$. 
	When $|\phi _{2,2}|=|\omega _{2,2}|$ holds, we obtain a quantity called the threshold resonance of $U$ as the quantity corresponding to the eigenvalue $\mu _0$ of $U$ obtained in the case $|\phi _{2,2}|<|\omega _{2,2}|$ in Theorem \ref{main2}.
	First, we define resonance and threshold resonance.

	\begin{definition}[Threshold resonances]
		We can naturally extend $U$ to $U_{\infty }$ acting on ${\mathcal H}_{\infty }:=\ell ^{\infty }(\mathbb {Z}^2;\mathbb {C} ^4)$. 
		Let $\Psi \in \mathcal{H}_{\infty}$ be a solution of $U_{\infty }\Psi =m \Psi $, where $m$ is a complex number.
		If $\Psi \in {\mathcal H}_{\infty }\backslash {\mathcal H}$, then we say that $m$ is a resonance of $U$ and $\Psi $ is a generalized eigenfunction of $U$ corresponding to $m$. 
		If $m$ is a resonance of $U$ and $m\in \partial \sigma _{\rm ess}(U)$ holds, then $m$ is called a threshold resonance. 
	\end{definition}
	
	\begin{remark}\label{rem:norm}
		Note that $\ell^2({\mathbb Z}^2;{\mathbb C}^4)\subsetneq \ell^{\infty }({\mathbb Z}^2;{\mathbb C}^4)$ and $\|\Psi \|_{\infty }\leq \|\Psi \|_{2}$ for any $\Psi \in \ell^2({\mathbb Z}^2;{\mathbb C}^4)$.
	\end{remark}
	
	We set $m_{\pm }:=e^{i\arccos (\pm \Lambda )}$. Under the conditons \eqref{c1}--\eqref{c3}, $\sigma _{\rm ess}(T)=[-\Lambda ,\Lambda ]$ holds by \eqref{eq:ess}. 
	Thus $m_{\pm}^\#\in \partial \sigma _{\rm ess}(U)$ by \eqref{SMT}, where $m_{\pm }^\#$ is $m_{\pm}$ or $m_{\pm}^*.$ 
	We use \eqref{eq:main2:1} to construct a candidate for the generalized eigenfunction of $U$ corresponding to $m_{\pm}^\#$. First, We set $\psi _{\pm \Lambda}({\bm x} ):=\lim _{\lambda \to \pm \Lambda \pm 0}\psi _{\lambda } ({\bm x} )$ for each ${\bm x} \in \mathbb {Z}^2$ 
	and 
	\begin{align*}
		\Sigma _{\pm}:=\{ e^{i\arccos \lambda } \mid  \lambda \in {\mathbb T}_{\pm}\},\quad 
		\Sigma _{\pm}^*:=\{ e^{-i\arccos \lambda } \mid  \lambda \in {\mathbb T}_{\pm}\}.
	\end{align*}
	Then $\bigcup _{\star =\pm }\left(\Sigma _{\star }\cup \Sigma _{\star}^*\right)={\mathbb S}^1\backslash \sigma _{\rm ess}(U)$ holds, where $\mathbb{S}^1$ is the unit circle on $\mathbb{C}$.
	Also, we define a vector 
	\begin{align}\label{eq:Psimu}
		\Psi _{\mu }:=(1-\mu S)d^*(\psi _{\lambda} -{\bm 1}_{\{\bm 0\}})\in \mathcal{H}=\ell^2(\mathbb{Z}^2;\mathbb{C}^4)
	\end{align}
	for $\mu \in {\mathbb S}^1\backslash \sigma _{\rm ess}(U)$ and $\lambda ={\rm Re}\, \mu \in {\mathbb T}_-\cup {\mathbb T}_+.$ 
	Then the following equations are satisfied in the same way as in the proof of Theorem~\ref{main2}:
	\begin{align}
		&\psi _{\lambda }({\bm x})={\bm 1}_{\{x_1=0\}}({\bm x})\int _{[0,2\pi)}\frac{2i\omega _{2,2}\phi _{2,1}\sin k_2}{2\phi _{2,1}\phi _{2,2}\cos k_2-\lambda }e^{ik_2x_2}\frac{dk_2}{2\pi},\label{eq:resopsi} \\
		&\Psi _{\mu }({\bm x})=\begin{pmatrix}
			\chi _{1,1}({\bm x})\left( \psi _{\lambda }({\bm x})-{\bm 1}_{\{0\}}({\bm x})\right)\\
			-\mu \chi _{1,1}({\bm x}-{\bm e}_1)\left( \psi _{\lambda }({\bm x}-{\bm e}_1)-{\bm 1}_{\{0\}}({\bm x}-{\bm e}_1)\right)\\
			\chi _{2,1}({\bm x})\left( \psi _{\lambda }({\bm x})-{\bm 1}_{\{0\}}({\bm x})\right)-\mu \chi _{2,2}({\bm x}+{\bm e}_2)\left( \psi _{\lambda }({\bm x}+{\bm e}_2)-{\bm 1}_{\{0\}}({\bm x}+{\bm e}_2)\right)\\
			\chi _{2,2}({\bm x})\left( \psi _{\lambda }({\bm x})-{\bm 1}_{\{0\}}({\bm x})\right)-\mu \chi _{2,1}({\bm x}-{\bm e}_2)\left( \psi _{\lambda }({\bm x}-{\bm e}_2)-{\bm 1}_{\{0\}}({\bm x}-{\bm e}_2)\right)\\
		\end{pmatrix},\label{eq:resoPsi}
	\end{align}   where $\mu \in {\mathbb S}^1\backslash \sigma _{\rm ess}(U)$ and $\lambda ={\rm Re}\, \mu \in {\mathbb T}_-\cup {\mathbb T}_+.$ 
	
	Also, we define 
	\begin{align*}
		\Psi _{m _{\pm}^\#}({\bm x} ):=\lim _{\stackrel{\mu \to m_{\pm}^\#}{\mu \in \Sigma _{\pm }^\#}}\Psi _{\mu}({\bm x} ),\quad {\bm x}\in {\mathbb Z}^2.
	\end{align*}
	
	The following are the main results of this subsection.
	\begin{theorem}\label{main3}
		Under the conditions \eqref{c1}--\eqref{c3}, we assume $|\phi _{2,2}|=|\omega _{2,2}|$. 
		Then $m_{\pm}^\#$ are threshold resonances of $U$, i.e., 
		\begin{align*}
			\Psi _{m_{\pm}^\#}\in {\mathcal H}_{\infty }\backslash {\mathcal H}, \
			U_{\infty }\Psi _{m_{\pm}^\#}=m_{\pm}^\#\Psi _{m_{\pm}^\#}.
		\end{align*}
	\end{theorem}
	
	\begin{proposition}\label{reso3}
		Under the conditions \eqref{c1}--\eqref{c3}, we assume $|\phi _{2,2}|=|\omega _{2,2}|$. Then the following equations hold for each $\bm{x}\in \mathbb{Z}^2$: 
		\begin{align}
			\psi _{\pm\Lambda}({\bm x})=(\pm {\rm sgn}(\phi _{2,1}\phi _{2,2}))^{x_2}{\rm sgn}(x_2)\frac{\omega _{2,2}}{\phi _{2,2}}\bm{1}_{\{x_1=0\}\cup \{x_2\not= 0\}}({\bm x}). \label{eq:reso3}
		\end{align}
		In particular, the above equations \eqref{eq:reso3} show that $\Psi _{m_{\pm}^\#}\in {\mathcal H}_{\infty }\backslash {\mathcal H}$ holds. 
	\end{proposition}
	
	\begin{proof}
		For $\lambda\in \mathbb{T}_-\cup \mathbb{T}_+$ and 
		$x\in \mathbb{Z}$, we set 
		\begin{align*}
			I_{\pm}(x)&:=\lim _{\lambda \to \pm\Lambda\pm 0}\int _{-\pi}^{\pi}\frac{-i\sin k}{2\phi _{2,1}\phi _{2,2}\cos k-\lambda }e^{ikx}\frac{dk}{2\pi} \\
			&= 
			2\lim _{\lambda \to  \pm\Lambda\pm 0}\int _0^{\pi}\frac{\sin k\sin xk}{2\phi _{2,1}\phi _{2,2}\cos k-\lambda }\frac{dk}{2\pi}.
		\end{align*}
		Then $\psi _{\pm \Lambda }({\bm x})=\lim_{\lambda\to \pm\Lambda\pm 0}\psi_{\lambda}(\bm{x}) = -2\omega _{2,2}\phi _{2,1}{\bm 1}_{\{x_1=0\}}({\bm x})I_{\pm}(x_2)$ by \eqref{eq:resopsi}.
		
		First, we consider the case where $\phi _{2,1}\phi _{2,2}>0$ holds. 
		In this case, since $\Lambda =2\phi _{2,1}\phi _{2,2}<\lambda $, we have
		\begin{align}
			\left|\frac{\sin k\sin xk}{2\phi _{2,1}\phi _{2,2}\cos k-\lambda }\right|\leq 
			\frac{1}{\Lambda}\frac{|\sin k\sin xk|}{1-\cos k}
			,\quad k\in [0,\pi],\ x\in \mathbb{Z}.\label{upper_eval}
		\end{align}
		The right-hand side of \eqref{upper_eval} is found to be bounded with respect to $k$ on $(0,\pi]$ by continuity and the following calculation: 
		\begin{align*}
			\lim _{k\to 0+0}\frac{|\sin k\sin xk|}{1-\cos k}&=\lim _{k\to 0+0}\frac{\sin xk}{\sin k}\frac{k}{xk}(1+\cos k)x=2x,\quad x\in {\mathbb Z}\backslash \{0\}.
		\end{align*}
		Thus, the following inequality holds for all $x\in {\mathbb Z}$: 
		\begin{align*}
			\int _0^{\pi}\frac{1}{\Lambda}\frac{|\sin k\sin xk|}{1-\cos k}\frac{dk}{\pi}<\infty .
		\end{align*}
		Therefore, by the dominated convergence theorem, we obtain 
		\begin{align*}
			I_{+}(x)=-\frac{1}{\Lambda \pi}\int _0^{\pi}\frac{\sin k\sin xk}{1-\cos k}dk
		\end{align*}in the case of $\phi _{2,1}\phi _{2,2}>0$. We set \begin{align*}
			J_{\pm}(x):=\int _0^{\pi}\frac{\sin k\sin xk}{1\pm \cos k}dk,\quad x\in \mathbb {Z},
		\end{align*}
		then $I_+(x)=-(1/\Lambda \pi)J_-(x)$ for $x\in {\mathbb Z}$ in the case of $\phi _{2,1}\phi _{2,2}>0$.
		
		In the other cases, we have the following table for $I_{\pm}$ in the same way.
		\begin{table}[htb]
			\begin{center}
				\begin{tabular}{|c|c|c|} \hline
					& $\phi _{2,1}\phi _{2,2}>0$  &  $\phi _{2,1}\phi _{2,2}<0$ \\ \hline 
					$I_+(x)$	& $-\frac{1}{\Lambda \pi}J_-(x)$ &   $-\frac{1}{\Lambda \pi}J_+(x)$  \\ \hline
					$I_-(x)$	&  $\frac{1}{\Lambda \pi}J_+(x)$ &   $\frac{1}{\Lambda \pi}J_-(x)$ \\ \hline
				\end{tabular}
			\end{center}
		\end{table}
		
		For $x\geq 4$, we have
		\begin{align*}
			J_{\pm}(x)
			&=\int _0^{\pi}\frac{\sin k\left( \sin k \cos(x-1)k+\cos k\sin (x-1)k\right)}{1\pm \cos k}dk\\
			&=\int _0^{\pi}\left(1\mp \cos k\right)\cos(x-1)k\ dk\\
			&\quad +\int _0^{\pi}\frac{\sin k\cos k}{1\pm \cos k}\left(\sin k\cos(x-2)k+\cos k\sin (x-2)k\right)dk\\
			&=\int _0^{\pi}\frac{\left( \sin k-\sin ^3k\right)\sin (x-2)k}{1\pm \cos k}dk\\
			&=J_{\pm}(x-2)\\
			&=\begin{cases}
				J_{\pm}(2),\quad x:\text{even}, \\
				J_{\pm}(3),\quad x:\text{odd}.
			\end{cases}
		\end{align*}
		Note that $J_{\pm}(-x) = - J_{\pm}(x)$ for all $x\in \mathbb{Z}$. 
		A direct calculation using these relational equations for $J_{\pm}(x)$ yields the following equations: 
		\begin{align*}
			J_{\pm }(x)=
			\begin{cases}
				{\rm sgn}(x)(\mp 1)^{x+1}\pi, & x \neq 0,\\
				0, & x = 0.
			\end{cases}
		\end{align*}
		Hence by \eqref{eq:resopsi}, we obtain \eqref{eq:reso3}.
		In particular, because $|\psi _{\pm \Lambda }({\bm x})|=\bm{1}_{\{x_1=0\}\cup \{x_2\not= 0\}}({\bm x})$ and \eqref{eq:resoPsi}, we get $\Psi _{m_{\pm}^\#}\in {\mathcal H}_{\infty }\backslash {\mathcal H}.$
	\end{proof}

	We set $\varepsilon (\lambda):=-\langle \varphi _{\bm{q}},\psi _{\lambda}\rangle /\lambda$ for all $\lambda \in \mathbb{T}_-\cup \mathbb{T}_+.$ 
	\begin{lemma}
		\label{lem:epsilon}We assume \eqref{c1}--\eqref{c3} and $|\phi _{2,2}|=|\omega _{2,2}|$. Then $\lim _{\lambda \to \pm \Lambda\pm 0}\varepsilon(\lambda)=1$ holds.
	\end{lemma}
	
	\begin{proof}
		By the assumption $|\phi _{2,2}|=|\omega _{2,2}|$ and the proof of Theorem~\ref{main1}, we have the following equation:
		\begin{align*}
			\frac{f(\lambda )}{2\phi _{2,1}\phi _{2,2}}={\rm sgn}(a)\sqrt{a^2-1},\quad \text{ for } \lambda \in \mathbb{T}_-\cup \mathbb{T}_+,
		\end{align*}
		where $a=\lambda /2\phi _{2,1}\phi _{2,2}$.
		This means that $\lim _{\lambda \to \pm \Lambda\pm 0}f(\lambda )=0$.
		Since $\varepsilon(\lambda )=-f(\lambda )/\lambda +1$ and $\pm \Lambda \not =0,$ we have $\lim _{\lambda \to \pm \Lambda\pm 0}\varepsilon(\lambda)=1.$
	\end{proof}
	
	\begin{proposition}\label{reso1}
		Under the conditions \eqref{c1}--\eqref{c3}, we assume $|\phi _{2,2}|=|\omega _{2,2}|$. For all $\mu \in {\mathbb S}^1\backslash \sigma _{\rm ess}(U)$, we set $\lambda :={\rm Re}\,  \mu\in \mathbb{T}_-\cup \mathbb{T}_+$. Then 
		the following equation holds:
		\begin{align*}
			(U-\mu )\Psi _{\mu }=(\varepsilon(\lambda )-1)\{ 2\mu Sd^*\varphi_{\bm q}+(U-\mu)(1-\mu S)d^*{\bm 1}_{\{ \bm{0}\}}\}.
		\end{align*}
		In particular, 
		\begin{align*}
			\lim _{\stackrel{\mu \to m_{\pm}^\#}{\mu \in \Sigma _{\pm }^\#}}\| (U-\mu )\Psi _{\mu}\|_2=0.
		\end{align*}
	\end{proposition}
	
	Before proving Proposition~\ref{reso1}, we prepare some related lemmas. 
	Following equation \eqref{proto_h}, we define $h_{\lambda}$ as follows: 
	\begin{align*}
		h_{\lambda}:=[P-(P^{\perp}(T-\lambda)P^{\perp})^{-1}P^{\perp}TP]\psi _{\lambda }.
	\end{align*}
	\begin{lemma}\label{lem1:reso1}
		$h_{\lambda}=\psi _{\lambda }-\varepsilon(\lambda ){\bm 1}_{\{{\bf 0}\}}$. 
	\end{lemma}
	\begin{proof}
		By the same argument as the proof of Theorem~\ref{main2}, we have $(P^{\perp}(T-\lambda )P^{\perp})^{-1}=-\lambda ^{-1}P^{\perp}$ and $P^{\perp}TP=|{\bm 1}_{\{{\bf 0}\}}\rangle \langle \varphi _{\bm q}|.$ Moreover, $\psi _{\lambda}\in {\rm Ran}P$ holds. Thus we obtain
		\begin{align*}
			h_{\lambda}=P\psi _{\lambda }-\lambda ^{-1}P^{\perp}TP
			\psi _{\lambda}=\psi _{\lambda }-\varepsilon(\lambda ){\bm 1}_{ \{{\bf 0}\}}.	\end{align*} 
	\end{proof}

	\begin{lemma}\label{lem2:reso1}
		$(T-\lambda )h_{\lambda}=(1-\varepsilon(\lambda ))\varphi _{\bm q}.$
	\end{lemma}
	
	\begin{proof}
		By \cite[Lemma~3.2]{FFS}, the discriminant operator $T$ can be written as follows: 
		\begin{align}\label{disc_ffs}
			T=\chi _{2,1}L_2\chi _{2,2}+\chi _{2,2}L_2^*\chi _{2,1}, 
		\end{align} 
		where $\chi _{2,1}$ and $\chi _{2,2}$ are multiplication operators by $\chi _{2,1}({\bm x})$ and $\chi _{2,2}({\bm x})$, respectively. Using equation \eqref{eq:t} and \eqref{disc_ffs}, we obtain
		\begin{align*}
			&(T_0h_{\lambda})({\bm x})=\phi _{2,1}\phi _{2,2}((L_2h_{\lambda}) ({\bm x})+(L_2^*h_{\lambda} )({\bm x})),\\
			&(Th_{\lambda})({\bm x})=\chi _{2,1}({\bm x})\chi _{2,2}({\bm x}+{\bm e}_2)(L_2h_{\lambda})({\bm x})+\chi _{2,2}({\bm x})\chi _{2,1}({\bm x}-{\bm e}_2)(L_2^*h_{\lambda})({\bm x}).
		\end{align*} We define an operator $W$ and a function $\beta $ as $W:=T-T_0$ and $\beta ({\bm x}):=\chi _{2,1}({\bm x})\chi _{2,2}({\bm x}+{\bm e}_2)-\phi _{2,1}\phi _{2,2}$. 
		Here, noting $\psi _{\lambda }\in {\rm Ran}P$, we have $\psi _{\lambda }(0)=0$. With this fact and Lemma~\ref{lem1:reso1}, we obtain 
		\begin{align}
			(Wh_{\lambda})({\bm x})&=\left ( \beta ({\bf 0}){\bm 1}_{\{{\bf 0}\}}({\bm x})+\beta (-{\bm e}_2){\bm 1}_{\{-{\bm e}_2\}}({\bm x}) \right)(L_2h_{\lambda})({\bm x})\notag\\
			&\quad +\left ( \beta (-{\bm e}_2){\bm 1}_{\{{\bf 0}\}}({\bm x})+\beta ({\bf 0}){\bm 1}_{\{{\bm e}_2\}}({\bm x}) \right)(L_2^*h_{\lambda})({\bm x})\notag\\
			&=\beta ({\bf 0})\psi _{\lambda }({\bm e}_2){\bm 1}_{\{{\bf 0}\}}({\bm x})-\varepsilon(\lambda )\beta (-{\bm e}_2){\bm 1}_{\{-{\bm e}_2 \}}({\bm x})\notag\\
			&\quad +\beta (-{\bm e}_2)\psi _{\lambda }(-{\bm e}_2){\bm 1}_{\{{\bf 0}\}}({\bm x})-\varepsilon(\lambda )\beta ({\bf 0}){\bm 1}_{\{ {\bm e}_2 \}}({\bm x}).\label{wh_lamda}
		\end{align}
		On the other hand, we get the following equation by \eqref{eq:phipsi} and Lemma~\ref{lem1:reso1}:
		\begin{align}
			(T_0-\lambda )h_{\lambda}=\varphi _{\bm q}-\varepsilon(\lambda )\phi _{2,1}\phi _{2,2}\left (  {\bm 1}_{ \{ -{\bm e}_2  \}}+ {\bm 1}_{ \{ {\bm e}_2  \}}\right) +\lambda \varepsilon(\lambda ){\bm 1}_{\{{\bf 0}\}}.\label{T0h_lambda}
		\end{align}
		Combining \eqref{wh_lamda} and \eqref{T0h_lambda}, we have
		\begin{align*}
			(T-\lambda )h_{\lambda}&=(T_0-\lambda )h_{\lambda}+Wh_{\lambda}\\
			&=\varphi _{\bm q}+
			A_1
			{\bm 1}_{\{{\bf 0}\}}
			-\varepsilon(\lambda )
			A_2
			{\bm 1}_{\{ -{\bm e}_2 \}}
			-\varepsilon(\lambda )
			A_3
			{\bm 1}_{\{ {\bm e}_2 \}}, 
		\end{align*}
		where
		\begin{align*}
			&A_1:=\beta ({\bf 0})\psi _{\lambda }({\bm e}_2)+\beta (-{\bm e}_2)\psi _{\lambda }(-{\bm e}_2)+\lambda \varepsilon(\lambda ),\\
			&A_2:=\beta (-{\bm e}_2)+\phi _{2,1}\phi _{2,2}, \\
			&A_3:=\beta ({\bf 0})+\phi _{2,1}\phi _{2,2}.
		\end{align*}
		By \eqref{eq:phipsi},
		\begin{align*}
			-\lambda \varepsilon(\lambda) &= 
			\langle \omega _{2,1}\phi _{2,2}{\bm 1}_{\{ {\bm e}_2 \}}+\omega _{2,2}\phi _{2,1}{\bm 1}_{ \{ -{\bm e}_2 \} },\psi _{\lambda }\rangle \\
			&=\omega_{2,1}\phi _{2,2}\psi _{\lambda }({\bm e}_2)+\omega_{2,2}\phi _{2,1}\psi _{\lambda }(-{\bm e}_2)\\
			&=\beta ({\bf 0})\psi _{\lambda }({\bm e}_2)+\beta (-{\bm e}_2)\psi _{\lambda }(-{\bm e}_2)+\phi _{2,1}\phi _{2,2}\left ( \psi _{\lambda }({\bm e}_2)+\psi _{\lambda }(-{\bm e}_2) \right).
		\end{align*}
		Since $2\cos k\cdot (-2i\omega _{2,1}\phi _{2,2}\sin k)/(2\phi _{2,1}\phi _{2,2}\cos k-\lambda )$ is an odd function, we obtain 
		\begin{align*}
			\psi _{\lambda }({\bm e}_2)+\psi _{\lambda }(-{\bm e}_2)=\int _{-\pi}^{\pi} 2\cos k_2\cdot \frac{-2i\omega _{2,1}\phi _{2,2}\sin k_2}{2\phi _{2,1}\phi _{2,2}\cos k_2-\lambda } \frac{dk_2}{2\pi}=0
		\end{align*}by \eqref{eq:resopsi}.
		Thus we get $A_1=\langle \varphi _{\bm q},\psi _{\lambda }\rangle +\lambda \varepsilon(\lambda )=0$.  
		On the other hand, we obtain $A_2=\omega _{2,2}\phi _{2,1},\ A_3=\omega _{2,1}\phi _{2,2}$ by the definition of $\beta .$
		Hence, we have
		\begin{align*}
			(T-\lambda )h_{\lambda}=(1-\varepsilon(\lambda ))\varphi _{\bm q}.
		\end{align*}
	\end{proof}

	\begin{proof}[Proof of Proposition \ref{reso1}]
		Lemma~\ref{lem1:reso1} and \eqref{eq:Psimu} lead to the following equation: 
		\begin{align*}
			\Psi _{\mu }=(1-\mu S)d^*h_{\lambda}+(\varepsilon(\lambda )-1)(1-\mu S)d^*{\bm 1}_{\{ \bm{0}\}}.
		\end{align*}
		Then we have
		\begin{align*}
			(U-\mu )\Psi _{\mu }&=(SC-\mu )(1-\mu S)d^*h_{\lambda}+(\varepsilon(\lambda )-1)(U-\mu)(1-\mu S)d^*{\bm 1}_{\{ \bm{0}\}}\\
			&=SCd^*h_{\lambda}-\mu d^*h_{\lambda}-\mu SCSd^*h_{\lambda}+\mu^2Sd^*h_{\lambda}+(\varepsilon(\lambda )-1)(U-\mu)(1-\mu S)d^*{\bm 1}_{\{ \bm{0}\}}\\
			&=
			(1-2\mu \lambda +\mu^2)Sd^*h_{\lambda}+2\mu (\varepsilon(\lambda )-1)Sd^*\varphi_{\bm q}+(\varepsilon(\lambda )-1)(U-\mu)(1-\mu S)d^*{\bm 1}_{\{ \bm{0}\}}. 
		\end{align*}
		The last equality in the above equation is due to the fact that 
		$SCd^*h_{\lambda}=Sd^*h_{\lambda},\ SCSd^*=S(2d^*d-1)Sd^*=2Sd^*T-d^*$ and Lemma~\ref{lem2:reso1}.
		Also, the equation $1-2\mu \lambda +\mu^2=0$ holds, since $\lambda ={\rm Re}\ \mu $ and $|\mu |=1.$ Thus
		\begin{align*}
			(U-\mu )\Psi _{\mu }=(\varepsilon(\lambda )-1)\{ 2\mu Sd^*\varphi_{\bm q}+(U-\mu)(1-\mu S)d^*{\bm 1}_{\{ \bm{0}\}}\}.
		\end{align*}
		Therefore, by $|\mu |=1$, Lemma~\ref{lem:epsilon} and unitarity of $U$ and $S$, we obtain 
		\begin{align*}
			\lim _{\stackrel{\mu \to m_{\pm}^\#}{\mu \in \Sigma _{\pm }^\#}}\| (U-\mu )\Psi _{\mu }\|_2&= \lim _{\lambda \to \pm \Lambda \pm 0}|\varepsilon(\lambda )-1|\| 2\mu Sd^*\varphi_{\bm q}+(U-\mu)(1-\mu S)d^*{\bm 1}_{\{ \bm{0}\}}\|_2\\
			&\leq \lim _{\lambda \to \pm \Lambda \pm 0}2|\varepsilon(\lambda )-1|\left( \| Sd^*\varphi_{\bm q}\|_2+2\|d^*{\bm 1}_{\{ \bm{0}\}}\|_2  \right)=0. 
		\end{align*}
	\end{proof}
	
	We are now ready to prove our main result on threshold resonances in this subsection, 
	Theorem~\ref{main3}.
	
	\begin{proof}[Proof of Theorem~\ref{main3}]
		In what follows, we show that $m_+$ is a threshold resonance of $U$.
		By the triangle inequality, for $\mu \in \Sigma _+$ and ${\bm x}\in {\mathbb Z}^2,$ we obtain 
		\begin{align}
			\| (U_{\infty }\Psi _{m_{+}})({\bm x} )-m_{+}\Psi_{m_{+}}({\bm x} )\|_{\mathbb {C} ^4}&\leq \| (U_{\infty }\Psi _{m_{+}})({\bm x} )-(U_{\infty }\Psi _{\mu})({\bm x} )\|_{\mathbb {C} ^4}\label{eq:reso.main.1}\\
			&+\| (U_{\infty }\Psi _{\mu})({\bm x} )-\mu \Psi _{\mu }({\bm x} ) \|_{\mathbb {C} ^4}\label{eq:reso.main.2}\\
			&+\| \mu \Psi _{\mu }({\bm x} ) -m_{+}\Psi _{m_{+}}({\bm x})\|_{\mathbb {C} ^4}.\label{eq:reso.main.3}
		\end{align}
		By the definition of $\Psi _{m_{+}}$, the term \eqref{eq:reso.main.3} converges to $0$ as $\mu \to m_{+}$.
		Noting Remark~\ref{rem:norm}, 
		the term \eqref{eq:reso.main.2} can be evaluated from above as follows: 
		\begin{align}
			\| (U_{\infty }\Psi _{\mu})({\bm x} )-\mu \Psi _{\mu }({\bm x} ) \|_{\mathbb {C} ^4}\leq \| U_{\infty }\Psi _{\mu}-\mu \Psi _{\mu } \|_{\infty }\leq \| U\Psi _{\mu}-\mu \Psi _{\mu } \|_{2}. \label{U_inf}
		\end{align}
		By Proposition~\ref{reso1}, the right-hand side of \eqref{U_inf} converges to $0$ as $\mu \to m_{+}.$ Then the term \eqref{eq:reso.main.2} converges to $0$ as $\mu \to m_{+}.$
		Finally, we prove that the term \eqref{eq:reso.main.1} converges to $0$ as $\mu \to m_+.$
		Let $\Phi  _{\mu }:=\Psi _{\mu }-\Psi _{m_+}$ 
		and $C_{\infty }$ be the natural extension of $C$ acting on ${\mathcal H}_{\infty }$. 
		We have
		\begin{align*}
			(U_{\infty }\Phi _{\mu })({\bm x} )=\begin{pmatrix}
				(C_{\infty }\Phi _{\mu })_2({\bm x} +{\bm e}_1)\\
				(C_{\infty }\Phi _{\mu })_1({\bm x} -{\bm e}_1)\\
				(C_{\infty }\Phi _{\mu })_4({\bm x} +{\bm e}_2)\\
				(C_{\infty }\Phi _{\mu })_3({\bm x} -{\bm e}_2)\\
			\end{pmatrix}, 
		\end{align*}
		here $(C_{\infty }\Phi _{\mu })_i({\bm x} )$ is the $i$-th component of $(C_{\infty }\Phi _{\mu })({\bm x} )$ for each ${\bm x} \in \mathbb {Z}^2$, $i=1,2,3,4$. 
		Since $C_{\infty }$ is one-defect, there exists a constant $c>0$ such that $|(C_{\infty }\Phi _{\mu })_i({\bm x} )|\leq c|(\Phi _{\mu})_i({\bm x} )|$ for ${\bm x} \in \mathbb {Z}^2,\ i=1,2,3,4.$ Thus we obtain 
		\begin{align*}
			\| (U_{\infty }\Phi _{\mu })({\bm x} )\|_{\mathbb {C} ^4}^2
			&\leq c\{ |(\Phi _{\mu })_1({\bm x} -{\bm e}_1)|^2+|(\Phi _{\mu })_2({\bm x} +{\bm e}_1)|^2+|(\Phi _{\mu })_3({\bm x} -{\bm e}_2)|^2+|(\Phi _{\mu })_4({\bm x} +{\bm e}_2)|^2  \} \\
			&\to 0
		\end{align*}
		as $\mu \to m_{+}.$ This means that the term \eqref{eq:reso.main.1} convergences to $0$ as $\mu \to m_+.$ 
		Hence $\| (U_{\infty }\Psi _{m_+})({\bm x} )-m_+\Psi_{m_+}({\bm x} )\|_{\mathbb {C} ^4}=0$ holds for all $\bm{x}\in\mathbb{Z}^2$, i.e., 
		\begin{align*}
			U_{\infty }\Psi _{m_{+}} =m_{+}\Psi_{m_{+}}. 
		\end{align*}
		In addition to the above results, by Proposition~\ref{reso3}, $m_+$ is a threshold resonance of $U$.
		For other threshold resonances, the theorem can be proved in the same way as above.
	\end{proof}

	\medskip
	
	\noindent
	{\bf Acknowledgement}
	This work was supported by JSPS KAKENHI Grant number 19K14596 (T.F.).


\begin{thebibliography}{10}
	
	\bibitem{ALMM}
	Sergio Albeverio, Saidakhmat~N Lakaev, Konstantin~A Makarov, and Zahriddin~I
	Muminov.
	\newblock The threshold effects for the two-particle hamiltonians on lattices.
	\newblock {\em Communications in mathematical physics}, 262(1):91--115, 2006.
	
	\bibitem{BFS}
	Volker Bach, J{\"u}rg Fr{\"o}hlich, and Israel~Michael Sigal.
	\newblock Renormalization group analysis of spectral problems in quantum field
	theory.
	\newblock {\em Advances in Mathematics}, 137(2):205--298, 1998.
	
	\bibitem{EKKT19}
	Takako Endo, Takashi Komatsu, Norio Konno, and Tomoyuki Terada.
	\newblock Stationary measure for three-state quantum walk.
	\newblock {\em Quantum Inf. Comput.}, 19(11$\&$12):901--912, 2019.
	
	\bibitem{FFS}
	Toru Fuda, Daiju Funakawa, and Akito Suzuki.
	\newblock Localization of a multi-dimensional quantum walk with one defect.
	\newblock {\em Quantum Information Processing}, 16(8):203, 2017.
	
	\bibitem{FFS2}
	Toru Fuda, Daiju Funakawa, and Akito Suzuki.
	\newblock Localization for a one-dimensional split-step quantum walk with bound
	states robust against perturbations.
	\newblock {\em Journal of Mathematical Physics}, 59(8):082201, 2018.
	
	\bibitem{FFS3}
	Toru Fuda, Daiju Funakawa, and Akito Suzuki.
	\newblock Weak limit theorem for a one-dimensional split-step quantum walk.
	\newblock {\em Rev. Roumaine Math. Pures Appl.}, 64:157--165, 2019.
	
	\bibitem{FNSS}
	Toru Fuda, Akihiro Narimatsu, Kei Saito, and Akito Suzuki.
	\newblock Spectral analysis for a multi-dimensional split-step quantum walk
	with a defect.
	\newblock {\em Quantum Studies: Mathematics and Foundations}, 9(1):93--112,
	2022.
	
	\bibitem{HL}
	Fumio Hiroshima and J{\'o}zsef L{\H{o}}rinczi.
	\newblock The spectrum of non-local discrete schr{\"o}dinger operators with a
	$\delta$-potential.
	\newblock {\em Pacific Journal of Mathematics for Industry}, 6(1):1--6, 2014.
	
	\bibitem{KKK}
	Hikari Kawai, Takashi Komatsu, and Norio Konno.
	\newblock Stationary measure for two-state space-inhomogeneous quantum walk in
	one dimension.
	\newblock {\em Yokohama Mathematical Journal}, 64:111--130, 2018.
	
	\bibitem{KBF}
	Takuya Kitagawa, Matthew~A Broome, Alessandro Fedrizzi, Mark~S Rudner, Erez
	Berg, Ivan Kassal, Al{\'a}n Aspuru-Guzik, Eugene Demler, and Andrew~G White.
	\newblock Observation of topologically protected bound states in photonic
	quantum walks.
	\newblock {\em Nature communications}, 3(1):1--7, 2012.
	
	\bibitem{Ki}
	Takuya Kitagawa, Mark~S Rudner, Erez Berg, and Eugene Demler.
	\newblock Exploring topological phases with quantum walks.
	\newblock {\em Physical Review A}, 82(3):033429, 2010.
	
	\bibitem{KK17}
	Takashi Komatsu and Norio Konno.
	\newblock Stationary amplitudes of quantum walks on the higher-dimensional
	integer lattice.
	\newblock {\em Quantum Information Processing}, 16(12):1--16, 2017.
	
	\bibitem{KK22}
	Takashi Komatsu and Norio Konno.
	\newblock Stationary measure induced by the eigenvalue problem of the
	one-dimensional hadamard walk.
	\newblock {\em Journal of Statistical Physics}, 187(1):1--24, 2022.
	
	\bibitem{KKMS}
	Takashi Komatsu, Norio Konno, Hisashi Morioka, and Etsuo Segawa.
	\newblock Generalized eigenfunctions for quantum walks via path counting
	approach.
	\newblock {\em Reviews in Mathematical Physics}, 33(06):2150019, 2021.
	
	\bibitem{KKMS2}
	Takashi Komatsu, Norio Konno, Hisashi Morioka, and Etsuo Segawa.
	\newblock Asymptotic properties of generalized eigenfunctions for
	multi-dimensional quantum walks.
	\newblock {\em Annales Henri Poincar{\'e}}, 23(5):1693--1724, 2022.
	
	\bibitem{LTRSW}
	Tania Loke, Judy~W Tang, Jeremy Rodriguez, Michael Small, and Jingbo~B Wang.
	\newblock Comparing classical and quantum pageranks.
	\newblock {\em Quantum information processing}, 16(1):1--22, 2017.
	
	\bibitem{MC}
	Arindam Mallick and CM~Chandrashekar.
	\newblock Dirac cellular automaton from split-step quantum walk.
	\newblock {\em Scientific reports}, 6(1):1--13, 2016.
	
	\bibitem{Ma}
	Yasumichi Matsuzawa.
	\newblock An index theorem for split-step quantum walks.
	\newblock {\em Quantum Information Processing}, 19(8):1--8, 2020.
	
	\bibitem{MST}
	Yasumichi Matsuzawa, Motoki Seki, and Yohei Tanaka.
	\newblock The bulk-edge correspondence for the split-step quantum walk on the
	one-dimensional integer lattice.
	\newblock {\em arXiv preprint arXiv:2105.06147}, 2021.
	
	\bibitem{Mo}
	Hisashi Morioka.
	\newblock Generalized eigenfunctions and scattering matrices for
	position-dependent quantum walks.
	\newblock {\em Reviews in Mathematical Physics}, 31(07):1950019, 2019.
	
	\bibitem{NOW}
	Akihiro Narimatsu, Hiromichi Ohno, and Kazuyuki Wada.
	\newblock Unitary equivalence classes of split-step quantum walks.
	\newblock {\em Quantum Information Processing}, 20(11):1--23, 2021.
	
	\bibitem{SG}
	Andreas Schreiber, Aur{\'e}l G{\'a}bris, Peter~P Rohde, Kaisa Laiho, Martin
	{\v{S}}tefa{\v{n}}{\'a}k, V{\'a}clav Poto{\v{c}}ek, Craig Hamilton, Igor Jex,
	and Christine Silberhorn.
	\newblock A 2d quantum walk simulation of two-particle dynamics.
	\newblock {\em Science}, 336(6077):55--58, 2012.
	
	\bibitem{SeSu}
	Etsuo Segawa and Akito Suzuki.
	\newblock Generator of an abstract quantum walk.
	\newblock {\em Quantum Studies: Mathematics and Foundations}, 3(1):11--30,
	2016.
	
	\bibitem{SS}
	Etsuo Segawa and Akito Suzuki.
	\newblock Spectral mapping theorem of an abstract quantum walk.
	\newblock {\em Quantum Information Processing}, 18(11):1--24, 2019.
	
	\bibitem{Si}
	Yutaka Shikano.
	\newblock From discrete time quantum walk to continuous time quantum walk in
	limit distribution.
	\newblock {\em Journal of Computational and Theoretical Nanoscience},
	10(7):1558--1570, 2013.
	
	\bibitem{S}
	Mario Szegedy.
	\newblock Quantum speed-up of markov chain based algorithms.
	\newblock In {\em 45th Annual IEEE symposium on foundations of computer
		science}, pages 32--41. IEEE, 2004.
	
	\bibitem{T}
	Yohei Tanaka.
	\newblock A constructive approach to topological invariants for one-dimensional
	strictly local operators.
	\newblock {\em Journal of Mathematical Analysis and Applications},
	500(1):125072, 2021.
	
\end{thebibliography}
\end{document}